\documentclass{iau} 
\usepackage{graphicx}

\def\draftversion{1} % 0 = clean     1 = draft       2 = referee (clean but with \new commands in bold)
\usepackage{amssymb,latexsym,graphicx,natbib,eufrak,times,amsmath,xspace,ifthen}

\ifthenelse{ \draftversion > 0 }{
} {
  
}

\ifthenelse{\equal{\draftversion}{1}}{
  \usepackage{xcolor}
  \newcommand{\sep}[1]{\par\begin{color}[rgb]{0,0.4,0}\begin{center}\hrule\end{center}\end{color}\par} % SEPARATOR: green horizontal line
  \newcommand{\todo}[1]{\begin{color}{red}\ \ifthenelse{\equal{#1}{}} {$\bullet\bullet\bullet$} {$\bullet$\ #1 $\bullet$}\end{color}} % TODO: red
  \newcommand{\idea}[1]{\begin{color}[rgb]{0,0.4,0}\textit{#1}\end{color}} % IDEA: green
  \newcommand{\sk}[1]{\begin{color}[rgb]{0.6,0,0.6}#1\end{color}} % SKIP: purple
  \newcommand{\toc}{\par\begin{color}[rgb]{0.6,0,0.6}\begin{center}\hrule\vspace{0.5mm}\begingroup\small\let\cleardoublepage\relax\let\clearpage\relax\mytoc\endgroup\vspace{0.5mm}\hrule\end{center}\end{color}\par} % TOC

  }{
  \newsavebox{\trashcan}

  \newcommand{\sep}[1]{}
  \newcommand{\todo}[1]{}
  \newcommand{\idea}[1]{}
  \newcommand{\sk}[1]{}
  \newcommand{\toc}{}

  }
 \makeatletter\newcommand\mytoc{\@starttoc{toc}}\makeatother % set TOC style
\long\def\symbolfootnote[#1]#2{\begingroup%
\def\thefootnote{\fnsymbol{footnote}}\footnote[#1]{#2}\endgroup} 

\newcommand{\bb}[1]{\ifmmode \mbox{\boldmath $ #1$} \else  \mbox{\boldmath $#1$} \fi}

\newcommand{\mmh}{\ensuremath{\textrm{\,--\,}}}    % long hyphen
\newcommand{\U}[1]{\ensuremath{\mathrm{~#1}}}     % units
\newcommand{\Myr}{\U{Myr}}          
\newcommand{\Gyr}{\U{Gyr}}          
\newcommand{\pc}{\U{pc}}

\newcommand{\Msun}{\U{M}_{\odot}}

\newcommand{\K}{\U{K}}

\newcommand{\lund}{Department of Astronomy and Theoretical Physics, Lund Observatory, Box 43, SE-221 00 Lund, Sweden}

\title[3 things they don't tell you about star clusters]{3 things they don't tell you about star clusters}

\author[Florent~Renaud]{Florent~Renaud}

\affiliation{\lund\\email: {\tt florent@astro.lu.se}}

\pubyear{2019}
\volume{351}  
\jname{Star Clusters: From the Milky Way to the Early Universe}
\editors{A. Bragaglia, M.B. Davies, A. Sills \& E. Vesperini, eds.}

\begin{document}
\maketitle

\begin{abstract}
Dense stellar systems in general and star clusters in particular have recently regained the interest of the extragalactic and even cosmology communities, due to the role they could play as actors and probes of re-ionization, galactic archeology and the dark matter content of galaxies, among many others. In the era of the exploitation and the preparation of large stellar surveys (Gaia, APOGEE, 4MOST, WEAVE), of the detection of gravitational waves mostly originating from dense regions like the cores of clusters (Ligo, LISA), and in an always more holistic view of galaxy formation (HARMONI, Euclid, LSST\footnote{Soon to be known as the Vera Rubin Survey Telescope (VRST).}), a complete theory on the formation and evolution of clusters is needed to interpret the on-going and forthcoming data avalanche. In this context, the community carries an effort to model the aspects of star cluster formation and evolution in galactic and even cosmological context. However, it is not always easy to understand the caveats and the shortcuts taken in theories and simulations, and their implications on the conclusions drawn. I take the opportunity of this document to highlight three of these topics and discuss why some shortcuts taken by the community are or could be misleading.
% \keywords{Keyword1, keyword2, keyword3, etc.}
\end{abstract}

%%%%%%%%%%%%%%%%%%%%%%%%%%%%%%%%%%%%%%%%%%%%%%%%%%%%%%%%
\firstsection % if your document starts with a section, remove some space above using this command.
\section{Young massive clusters are not local analogues of young globular clusters}

More precisely, it would be surprising that they were, but we still don't know. Number of authors claim that present-day young massive clusters (YMCs, formerly known as super star clusters) are local analogues of the globular clusters at the epoch of their formation in the early Universe. Although appealing, this point has not been proven either observationally or theoretically, and an increasing number of hints suggest otherwise. The temptation to draw a direct connection between YMCs and young globulars comes from the lack of observational (and theoretical) knowledge on the physics of cluster formation at high redshift ($z \gtrsim 2\mmh6$, $\gtrsim 10\mmh 13 \Gyr$ ago). As massive and dense stellar systems, YMCs could indeed, in principle, be used as local analogues to probe similar mechanisms.

Many YMCs are found in special conditions like interacting galaxies (e.g. the Antennae) that could resemble that of the denser Universe at high redshift (with the caveats listed below), but others exist in more common, quieter areas \citep{Portegies2010}, where a special role of the galactic environment is yet to be identified (e.g. 30 Doradus in the Large Magellanic Cloud). At the formation epoch of globulars, extreme conditions are ubiquitous due to the increased role of gas instabilities and the constant bombardment of satellite galaxies, such that the quiet environment might not even exist. This already suggests that not \emph{all} YMCs are analogues of young globular clusters, or in other words, that massive clusters form through a range of channels and that not all these channels are equally active at high and low redshift.

The cluster mass function (CMF) of YMCs is observed to be a power-law, compatible with a (debated) exponential truncation of the Schechter form \citep[see][]{Adamo2017}. This power-law connects to the properties of the turbulence in the interstellar medium \citep{Elmegreen2006}. If YMCs were analogues of young globulars, this would also be the initial CMF of globulars, and thus the transition to the present-day globular CMF (GCMF) would solely result from evolutionary processes. However, the GCMF is observed to be universal \citep[see][and references therein]{Vesperini2001}. Because the evolution varies with the galactic environment (e.g. through a diversity of tidal fields), it must have a environment-dependent imprint on the CMF. The only logical conclusion is thus that the initial CMF is also environment-dependent (instead of the power-law or the Schechter function of YMCs), and that the initial and evolutionary dependences balance toward an equilibrium state which is the universal GCMF \citep[e.g.][]{Vesperini1998, Parmentier2007}. 

Reasons why the initial CMF may not be universal are also why the formation process of globulars may differ from that of YMCs. Notable differences exist between the properties of the host galaxies at low and high redshift and it is not yet clear to what extent they could affect the formation of the massive clusters. A non-exhaustive list is given below.

\begin{itemize}
	\item Instabilities and turbulence. At the epoch of formation of globulars, the baryonic content of galaxies is dominated by gas (as opposed to $\sim 10\%$ gas fraction in the present-day Milky Way). Because of the dissipative nature of gas, the instabilities driving the formation of structures and, in turn, of the formation sites of massive clusters, are regulated by different criteria as in star-dominated galaxies \citep[see][in the context of disks]{Romeo2017,Romeo2018,Romeo2019}. For instance, this is observed as the transition from massive gas clumps in disks with a $\sim 50 \mmh60\%$ gas fraction to spiral arms in more stellar-dominated systems \citep{Agertz2009}, which results in different environments for the star (cluster) forming regions (clump mass, density, turbulence, shear, tides, see also \citealt{Dessauges2018}). Furthermore, the nature of turbulence changes during galactic interactions towards compression-dominated modes \citep{Renaud2014b}. Therefore, the structure of the ISM and thus the CMF likely differs between massive clumps (in gas-rich but isolated galaxies) and mergers (at low or high redshift), calling for more than one formation channel and an evolution of their relative important along cosmic time.
	\item Gas metallicity. At high redshift, all galaxies are metal-poor and cooling proceeds differently as in the local Universe, which modifies the thermal support of structures against collapse. This differences thus affect the hierarchical organization of the ISM and its fragmentation into the formation sites of clusters. Furthermore, the stellar winds (of which energy depends on the metallicity) also alter the structure of the gas while the cluster still forms (i.e. before the onset of supernovae). Although winds are not the most energetic form of feedback, together with other pre-SN effects (e.g. cosmic rays, photo-ionization, radiative pressure), they play a crucial role in setting the low-density high-temperature cavities in which the SN blast eventually expand, which is critical for the propagation of feedback effects and their influence on the cluster and/or nearby clusters \citep{Hayward2017, Ohlin2019}.
	\item Stellar initial mass function (IMF). Although the IMF seems to be universal in the solar neighborhood, number of studies have reported variations of its shape. Empirically, the galactic environment plays an important role, as the clearest variations of the IMF are found in massive elliptical galaxies, in dwarfs and at the tip of the Milky Way bar \citep{Cappellari2012, Geha2013, Motte2018}. Interestingly, \citet{Dib2018} showed that IMF variations from cluster to cluster could still average out into a universal IMF at galactic scale. An holistic understanding of the IMF is still lacking, but it is very likely that the IMF varies significantly from its canonical shape in the extreme conditions needed for massive cluster formation, in particular in the early Universe. If true, the fragmentation into pre-stellar cores, the mass segregation, the initial and evolved binarity, the injection of feedback, and the cluster mass-loss due to stellar evolution would also varies between YMCs and young globulars.
\end{itemize}

Other considerations (population III stars, formation in mini-halos, the effect of re-ionization etc.) also points toward differences in the conditions of massive cluster formation between high and low redshift. In the end, because the massive clusters of a galaxy like the Milky Way originate from many different environments (variation with cosmic time and accretion from several satellite galaxies, \citealt{Renaud2017}), this diversity could imprint the galactic-wide statistical properties of clusters and blur possible signatures of the formation stage.

The presence of multiple stellar populations detected in relatively young massive clusters ($\lesssim 6  \Gyr$, \citealt{Krause2016}) suggests that the (still unidentified) process(es) responsible for the formation of multiple populations do not significantly vary with redshift (see however Milone et al., this volume, for arguments that the physics of multiple populations is not the same in young and old clusters). But without a clear understanding of this process(es), it is too early to claim that this fact proves an analogy between YMCs and young globulars.

Finally, even if massive cluster formation proceeds the same way at low and high redshift, the differences in the dynamics of their hosts galaxies imply different early evolution, for instance in the hierarchical build-up of clusters via cluster-cluster collisions. This process is naturally evoked in the formation of nuclear star clusters and ultra compact dwarf galaxies \citep[e.g.][]{Fellhauer2005}, and the fuzzy boundary between these objects and globular clusters \citep{Renaud2018b} indicates that collisions could play a role in the assembly of the most massive clusters \citep{Fellhauer2003}. In a gas-rich galaxy for instance, a massive clump ($\sim 10^9 \Msun$) can host the formation of more than one cluster that could merge within a crossing time ($\sim 10 \Myr$). This situation is much rarer in the local Universe where it is only found in very specific locations (galactic centers, mergers). Thus, it remains to be shown, from first principles, whether the evolution of these special conditions only accounts for the rarity of YMCs, or if it indicates that some formation channel(s) (among others) becomes less active with cosmic time.

Because of our partial understanding of star formation physics in the early Universe, models and simulations of globular cluster formation usually \emph{assume} that the conditions for the formation of globulars is the same as for local YMCs\footnote{These conditions are still to be fully determined, but likely encompass cold, dense, turbulent and eventually self-gravitating media with large-scale contributions (shocks, convergent flows, compressive tides) and with weak destructive effects (shear, tides). The details likely varies in different environments within galaxies, from one galaxy to the next, and along the evolution of the cloud and star formation processes themselves.}. A detailed treatment of cluster evolution across cosmic time, with a predictive power, could then be used to test the formation hypothesis (but see the next section).

In summary, variations in the star and star cluster formation processes exist between the formation epoch of globulars and the local Universe where YMCs are detected, and it has not yet been established to what extent these variations have an impact on the resulting clusters. Furthermore, the present-day mass function of globulars seems incompatible with that of YMCs with our current (but incomplete) understanding of cluster evolution. Therefore, the claim that the physics of formation of YMCs and globulars is the same remains highly questionable.

%%%%%%%%%%%%%%%%%%%%%%%%%%%%%%%%%%%%%%%%%%%%%%%%%%%%%%%%
\section{We do not capture cluster mass-loss and dissolution}

Resolving the internal evolution of clusters, ruled by star-star interactions, remains out of reach of galaxy and cosmology simulations due to the large range of space- and time-scales involved (AU\mmh Mpc, day\mmh Gyr). Models for cluster evolution could be adopted to estimate the evolution of their mass, size and energy. They would require a description of the rapidly time-varying tidal field the clusters experience. This tidal evolution comprises two main aspects. First are the secular, adiabatically varying tides generated by large-scale structures like the host galaxy itself. This component is reasonably well captured in galaxy simulations, as long as the morphological structures inducing the large-scale and slowly evolving nature of the gravitational potential are resolved by a few elements (particles, cells or both). Typical galactic scalelenghts are of the order of 1 kpc (e.g. bulge, disk, halo), and modern zoom-in cosmological simulations now routinely capture such structures with ten resolution elements or more. The associated term in the tidal field is thus well resolved, even in cosmological context (see e.g. \citealt{Renaud2017, Li2018}).

The other component is the rapidly changing, shock-like contributions of gravitational interactions of the clusters with nearby structures on timescales shorter than the relaxation (or the crossing) time. This occurs when the orbit of a cluster crosses a galactic disk \citep[e.g.][]{Vesperini1997}, or when it interacts with dense gas structures like molecular clouds and/or sub-structures in molecular gas \citep[which is most frequent in gas-rich galaxies in the early Universe]{Elmegreen2010}. During such events the stars are rapidly tidally accelerated, such that they possibly escape the cluster before the cluster tidal energy decreases back (or in other words before the tidal boundary retrieves a large, pre-shock-like, size). The high amounts of energy injected into the cluster during shock-like events imply that neglecting this component could lead to important errors on the mass-loss and overall survival of clusters (e.g. \citealt{Gnedin1999}, but see also \citealt{Gieles2016} on the milder effect of repeated shocks). With incomplete information on the small-scale structures ($< 10 \pc$) making these rapid changes in the tidal field, the dynamical evolution of clusters cannot be inferred. Present-day cosmological simulations do not have sufficient resolution to capture these structures, and thus the tidal shocks.

For instance, the E-MOSAICS project \citep{Pfeffer2018, Kruijssen2019} claims capturing tidal shocks while the resolution of their simulations is several $100 \pc$. When computing the tidal field, these authors derive the potential at scales 200 times \emph{smaller} than resolution, and interpret the result as a succession of shocks. Any quantity measured with less than a few resolution elements (and all the more below resolution limit, like in E-MOASAIC) is unphysical, and likely relates to numerical noise induced by the derivation\footnote{These authors report convergence on the resulting mass-loss (not on the tidal field itself) when measuring tides over a range of scales, from 1/200 to 1 times the resolution. This further demonstrates that the effect of the shocks are not captured, since resolving smaller structures would add their contribution to net tidal field.} (as shown in \citealt{Renaud2010}). These rapid variations of the tides are not physical because the structures that would generate shocks (disk crossing, $\sim 100 \pc$, nearby clouds, $\sim 10 \pc$, filamentary structures within GMCs, $\sim 0.1 \pc$) are not resolved. In addition to the mere spatial resolution, another problem is the gas pressure of the warm ISM which supports gas structures and prevents their collapse into dense objects that could cause shocks. It is thus necessary to resolve the cold ($\lesssim 100 \K$) and small ($\lesssim 10 \pc$) features of the ISM to qualify for capturing tidal shocks. Currently, no cosmological simulation (not even in zoom-ins) meet these criteria.

In this context, the efforts of \citet{Li2017} stand apart: reaching $\sim 10 \pc$ resolution and incorporating a sub-grid model for turbulence, this work does describe \emph{some} components of the tidal shocks, induced by e.g. disk crossing and interactions with nearby clouds. The finest structures are still not captured and thus the mass-loss of clusters is still not fully described. However, this work further confirmed the importance of the small-scale, rapidly varying tidal component in altering the mass-loss of clusters and thus in setting the evolution of the CMF.

In conclusion, to date no cosmological simulation captures the tidal field with sufficient precision to predict the mass-loss of clusters. The only way forward is increasing the resolution of such simulations, in particular in the gaseous component, to describe scales comparable to that of clusters, i.e. $\sim 1 \pc$, and this with several resolution elements, i.e. a effective sub-parsec resolution in cosmological context. It is needless to say that such an increase in resolution \emph{must} be accompanied by a corresponding improvement of the sub-grid recipes for all the related physical mechanisms at such scales (e.g. a star-by-star description of feedback). As a consequence, the evolution from the initial CMF (still unknown, see previous section) to the present-day CMF is not yet understood and remains out of reach of the current generation of simulations.

%%%%%%%%%%%%%%%%%%%%%%%%%%%%%%%%%%%%%%%%%%%%%%%%%%%%%%%%
\section{The cluster formation rate does not correlate with the star formation rate}

All stars are thought to form in clusters \citep{Lada2003} because the conditions needed for the gas to become dense enough require larger than parsec-scale mechanisms to assemble the formation sites and keep them bound. An unstable region is thus large and massive enough to host the formation of more than one star. This does not imply that all clusters remain bound (which relates to the question of cluster formation efficiency, not addressed here). A similar but different topic is to know whether all stars form in new clusters, or if some form in existing clusters. 

The detection of multiple stellar populations in massive globular indicates that a given cluster can host the formation of several generations of stars, if the differences between the populations translate into differences in age (as suggested e.g. by \citealt{Bekki2017}). Such clusters then have an extended star formation history, with the second and subsequent generations contributing to the SFR but not to the cluster formation rate (CFR), as the latter would have become zero after the formation of the first generation. In that case, the SFR does not correlate with the CFR. 

Other cases of star formation in pre-existing clusters (and thus of mismatch between the CFR and the SFR) concern the assembly of nuclear star clusters by the accretion of gas in the existing central cluster of a galaxy \citep{Milosavljevic2004}\footnote{Note that another scenario proposes that nuclear clusters form by migration of gas-free clusters toward the center \citep{Tremaine1975}, and several works argue for a mix of the two formation channels \citep{Feldmeier2014, denbrok2014}.}. Such accretion fuels the central star forming region with gas (and possibly with stars too, see \citealt{Guillard2016}). This mechanism is amplified during galaxy mergers (and in barred galaxies to a lower extent) where kpc-scale torques fuel the matter inward (inside co-rotation, \citealt{Keel1985}). This mechanism participates in the build-up of complex populations in galactic centers \citep{Renaud2019b}, but is also be relevant for non-central clusters. If formed in a satellite galaxy, a nuclear cluster could be dense enough to survive the tidal stripping of its host during the accretion into a massive galaxy (e.g. the Milky Way). The remnant would then resemble a globular cluster in the halo of its new host (or, depending on its initial mass and mass-loss, to an ultra compact dwarf, see \citealt{Pfeffer2013}). However, the fraction of the Milky Way's globular clusters which have followed this route is still unknown.

In the previous example, the existing nuclear cluster constitutes a convergence point in the gas flows. Although this is obviously greatly facilitated by the background potential in galactic centers, it is possible that some clusters are dense and massive enough to attract surrounding gas by themselves, with the possible help of large-scale effects like converging flows (e.g. spiral arms, shocks in mergers), compressive tides, low shear and low destructive tides, i.e. in extreme conditions \citep{Peterson2009}. The relative role of these mechanisms in various environments is yet to be established, but simulations suggest that the CFR significantly deviates from the SFR at different degrees at different stages of a galactic interaction \citep[see e.g. Fig. 11 of][]{Renaud2015}.

Comparable situations are expected in the turbulent ISM of high redshift galaxies experiencing repeated interactions, during the formation of the bulk of globular clusters. It is thus possible that these extreme conditions could lead to an important mismatch between the CFR and the SFR, before the two quantities better trace each other in quieter environments like the present-day Milky Way. Here again, quantifying the importance of these mechanisms is necessary before reaching definite conclusions on the formation of star clusters.

%%%%%%%%%%%%%%%%%%%%%%%%%%%%%%%%%%%%%%%%%%%%%%

\emph{Acknowledgments:}
I thank the LOC and SOC for a interesting and flawless meeting, and for their invitation. I am indebted to Genevi\`eve Parmentier, and to many participants of the IAU 351 as well as to Oscar Agertz and Eric Andersson for interesting discussions and for encouraging me to write this paper. I thank Douglas Heggie and Ha-Joon Chang for the inspiration for the title of this contribution. I acknowledge support from the Knut and Alice Wallenberg Foundation.

%%%%%%%%%%%%%%%%%%%%%%%%%%%%%%%%%%%%%%%%%%%%%%%%%%%%%%%%
%% Bibliography
\newcommand{\aj}{AJ}% Astronomical Journal
\newcommand{\actaa}{Acta Astron.}% Acta Astronomica
\newcommand{\araa}{ARA\&A}% Annual Review of Astron and Astrophys
\newcommand{\apj}{ApJ}% Astrophysical Journal
\newcommand{\apjl}{ApJ}% Astrophysical Journal, Letters
\newcommand{\apjs}{ApJS}% Astrophysical Journal, Supplement
\newcommand{\ao}{Appl.~Opt.}% Applied Optics
\newcommand{\apss}{Ap\&SS}% Astrophysics and Space Science
\newcommand{\aap}{A\&A}% Astronomy and Astrophysics
\newcommand{\aapr}{A\&A~Rev.}% Astronomy and Astrophysics Reviews
\newcommand{\aaps}{A\&AS}% Astronomy and Astrophysics, Supplement
\newcommand{\azh}{AZh}% Astronomicheskii Zhurnal
\newcommand{\baas}{BAAS}% Bulletin of the AAS
\newcommand{\caa}{Chinese Astron. Astrophys.}% Chinese Astronomy and Astrophysics
\newcommand{\cjaa}{Chinese J. Astron. Astrophys.}% Chinese Journal of Astronomy and Astrophysics
\newcommand{\icarus}{Icarus}% Icarus
\newcommand{\jcap}{J. Cosmology Astropart. Phys.}% Journal of Cosmology and Astroparticle Physics
\newcommand{\jrasc}{JRASC}% Journal of the RAS of Canada
\newcommand{\memras}{MmRAS}% Memoirs of the RAS
\newcommand{\mnras}{MNRAS}% Monthly Notices of the RAS
\newcommand{\na}{New A}% New Astronomy
\newcommand{\nar}{New A Rev.}% New Astronomy Review
\newcommand{\pra}{Phys.~Rev.~A}% Physical Review A: General Physics
\newcommand{\prb}{Phys.~Rev.~B}% Physical Review B: Solid State
\newcommand{\prc}{Phys.~Rev.~C}% Physical Review C
\newcommand{\prd}{Phys.~Rev.~D}% Physical Review D
\newcommand{\pre}{Phys.~Rev.~E}% Physical Review E
\newcommand{\prl}{Phys.~Rev.~Lett.}% Physical Review Letters
\newcommand{\pasa}{PASA}% Publications of the Astron. Soc. of Australia
\newcommand{\pasp}{PASP}% Publications of the ASP
\newcommand{\pasj}{PASJ}% Publications of the ASJ
\newcommand{\qjras}{QJRAS}% Quarterly Journal of the RAS
\newcommand{\rmxaa}{Rev. Mexicana Astron. Astrofis.}% Revista Mexicana de Astronomia y Astrofisica
\newcommand{\skytel}{S\&T}% Sky and Telescope
\newcommand{\solphys}{Sol.~Phys.}% Solar Physics
\newcommand{\sovast}{Soviet~Ast.}% Soviet Astronomy
\newcommand{\ssr}{Space~Sci.~Rev.}% Space Science Reviews
\newcommand{\zap}{ZAp}% Zeitschrift fuer Astrophysik
\newcommand{\nat}{Nature}% Nature
\newcommand{\iaucirc}{IAU~Circ.}% IAU Cirulars
\newcommand{\aplett}{Astrophys.~Lett.}% Astrophysics Letters and Communications
\newcommand{\apspr}{Astrophys.~Space~Phys.~Res.}% Astrophysics Space Physics Research
\newcommand{\bain}{Bull.~Astron.~Inst.~Netherlands}% Bulletin Astronomical Institute of the Netherlands
\newcommand{\fcp}{Fund.~Cosmic~Phys.}% Fundamental Cosmic Physics
\newcommand{\gca}{Geochim.~Cosmochim.~Acta}% Geochimica Cosmochimica Acta
\newcommand{\grl}{Geophys.~Res.~Lett.}% Geophysics Research Letters
\newcommand{\jcp}{J.~Chem.~Phys.}% Journal of Chemical Physics
\newcommand{\jgr}{J.~Geophys.~Res.}% Journal of Geophysical Research
\newcommand{\jqsrt}{J.~Quant.~Spec.~Radiat.~Transf.}% Journal of Quantitiative Spectroscopy and Radiative Trasfer
\newcommand{\memsai}{Mem.~Soc.~Astron.~Italiana}% Mem. Societa Astronomica Italiana
\newcommand{\nphysa}{Nucl.~Phys.~A}% Nuclear Physics A
\newcommand{\physrep}{Phys.~Rep.}% Physics Reports
\newcommand{\physscr}{Phys.~Scr}% Physica Scripta
\newcommand{\planss}{Planet.~Space~Sci.}% Planetary Space Science
\newcommand{\procspie}{Proc.~SPIE}% Proceedings of the SPIE

\bibliographystyle{aa}

\end{document}